\documentstyle[aps,pra,multicol,epsf,array]{revtex}
\newcommand{\dket}[1]{ \left| #1 \right\rangle }
\newcommand{\dbra}[1]{ \left\langle #1 \right| }

\newcommand{\ba}[1]{\begin{array}{#1} }
\newcommand{\ea}{\end{array}}
\newcommand{\rt}{ \frac{1}{\sqrt{2}} }

\begin{document}
\title{Resource reduction via repeaters in entanglement distribution}
\author{A. Hutton and S. Bose}
\address{Centre for Quantum Computation, Clarendon Laboratory,
    University of Oxford,
    Parks Road,
    Oxford OX1 3PU, England}
\maketitle

\begin{abstract}
We show that the amount of entanglement needed as an initial
resource to set up a certain final amount of entanglement between
two ends of a noisy channel can be reduced in certain cases by
using quantum repeaters. Our investigation (for various channels)
considers cases when a large number of entangled pairs are
transmitted through the channel using known asymptotic results and
conjectured bounds on distillable entanglement.
\end{abstract}

\begin{multicols}{2}


\section{Introduction}

The resource of entanglement \cite{revs1,revs2,revs3} has many
useful applications in quantum information processing, such as
secret key distribution \cite{Ek}, teleportation \cite{ben} and
dense coding \cite{wies}. Recently, there has been an intense
interest in the physical implementation of these quantum
communication protocols \cite{impl}. In the future, it is
conceivable that a large number of distant users would want to
engage in communicating with each other through quantum protocols
for purposes of security \cite{Ek,ben84}, doubled capacity
\cite{wies}, high fidelity state transfer (by teleportation
\cite{ben}), reduced communication complexity \cite{cleve},
secret sharing \cite{secsha}, linking distant quantum computers
\cite{qcomp}, for distributed computation \cite{qdistr1,qdistr2}
and a host of other applications \cite{teleclone,pract}. For this
to happen, any pair or group of distant users will need to share
particles in any desired pure entangled state, irrespective of
noise in the entanglement distribution channels. In this context,
various general schemes which could directly or indirectly aid in
the distribution of entanglement have been theoretically studied
\cite{qdistr1,purf1,purf2,purf21,purf3,bhm,zuk,multswp,seriesswp,qreapt}
and experimentally demonstrated \cite{distrexpts}. One of these
schemes, suggested by Briegel {\em et al.} and D\"{u}r {\em et
al.}, is the idea of {\em quantum repeaters} \cite{qreapt}. In
their scheme, a number of nodes called repeaters are placed
between the ends of a noisy channel. The entanglement is first
distributed between neighboring nodes and concentrated by local
actions \cite{purf1,purf2,purf21,purf3} at these nodes to produce
a few highly entangled pairs. Then entanglement swapping
\cite{zuk,multswp,seriesswp} is used to connect the highly
entangled pairs in series to obtain some highly entangled pairs
between the ends of the noisy channel. In this way, the
exponential degradation of entanglement with channel length can
be prevented at the expense of only a logarithmic increase in
resources (cost) and a polynomial increase in time \cite{qreapt}.
In this paper, we will look at the use of repeaters from a rather
different angle. We will show that if the amount of entanglement
finally required between two ends of a noisy channel was fixed,
it could be achieved, in certain circumstances, at a much lower
cost using quantum repeaters.

\begin{figure}
\begin{center}
\leavevmode \epsfxsize=10cm \epsfbox{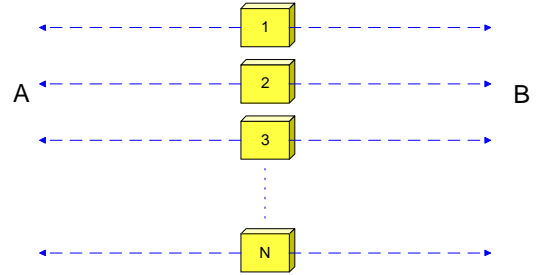} \vspace*{-7cm}
\caption{The figure shows a specific way of distributing
entanglement between two ends of a noisy channel. $N$ sources are
placed at the midpoint of the channel and made to emit maximally
entangled qubit pairs.} \label{fig1a}
\end{center}
\end{figure}
\vspace*{-1.0cm}
\begin{figure}
\begin{center}
 \leavevmode \epsfxsize=10cm
\epsfbox{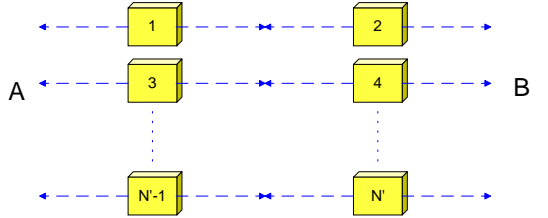} \vspace*{-10cm} \caption{The figure shows a
 way of distributing entanglement between two ends of a noisy channel which uses
 a quantum repeater positioned at the midpoint of the channel. With this configuration,
 $N^{'}<N$ sources of entangled pairs may be sufficient to establish as much entanglement between the
 ends of the channel as one without repeaters and $N$ sources.} \label{fig1b}
\end{center}
\end{figure}

  Consider the following scenario. You are given $N$ boxes labeled
$1$,$2$,...,$N$, each of which can be made to emit exactly one
maximally entangled particle pair on demand. Your target is to
distribute entanglement between two ends $A$ and $B$ of a noisy
channel. You have complete freedom of placing the boxes anywhere
along the length of the channel prior to distribution. Suppose
you first place all the $N$ sources at the midpoint of the channel
as shown in Fig.\ref{fig1a} and then make them emit maximally
entangled pairs. The pairs will lose part of their initial
entanglement while traveling from the middle to the end of the
channel. At the end of this process, a certain amount of
entanglement is obtained between the ends $A$ and $B$ of the
channel. You are then asked whether it is possible to devise a
more effective scheme in which you distribute the {\em same
amount of entanglement between A and B using a lower number of
initial sources of maximally entangled pairs}. It turns out, as
we will show, that the scheme shown in Fig.\ref{fig1b} can
accomplish this in certain circumstances ({\em i.e.} for certain
lengths and types of the noisy channels). The basic idea is to
divide the channel into two equal segments and place a repeater
station at the midpoint of the channel. A lower number $N^{'}$ of
sources may now be sufficient to distribute as much entanglement
between $A$ and $B$ as before. We first have to place $N^{'}/2$
sources at the midpoint of each half-channel. This is shown in
Fig.\ref{fig1b} with boxes $1$,$3$,...,$N^{'}-1$ having been
placed at the midpoint of the left half-channel and boxes
$2$,$4$,...,$N^{'}$ placed at the midpoint of the other
half-channel, where $N^{'}<N$. Then they are made to emit
maximally entangled pairs which are stopped at $A$, $B$ and the
repeater station.  After traversing a certain distance, each of
the pairs would have lost part of their entanglement. An operator
at the repeater station has to now act together with operators at
$A$ and $B$ to locally distill two sets of maximally entangled
pairs \cite{purf2}: one set between the repeater and $A$, and the
other set between the repeater and $B$. The operator at the
repeater then connects these maximally entangled pairs by
entanglement swapping \cite{zuk,multswp,seriesswp} to obtain a
set of maximally entangled states between $A$ and $B$.  Though
the maximum number of possible (entangled or disentangled)
particle pairs between $A$ and $B$ is $N^{'}/2 < N$ in the case
of distribution with a repeater, the degradation of entanglement
of each pair is also less (as each particle now travels half the
distance as before). The scheme with a repeaters saves resources
when the positive effect of lower entanglement loss on traversal
of the channel overrides the negative effect of having a lower
number of entanglement sources to start with.

    The resource reduction process described above, as we shall
demonstrate, can be made more significant for certain channel
lengths by dividing the channel into an even larger number of
segments and placing repeater stations at the junctions of these
segments. This would lead to an important {\em cost reduction} in
the distribution of entanglement. The cost of distributing
entanglement will become a very important issue in the future if
quantum communications take off and this topic has already
received serious attention (see, for example, Cirac {\em et al.}
\cite{qdistr1}). In this paper, we consider the cost reduction to
be approximately proportional to reduction of the initial
resource of entanglement (in terms of the number of sources of
maximally entangled pairs used). We neglect, for example, the cost
of classical communication during the entanglement distillation
procedures (though this cost might not necessarily increase on
using repeaters, as the number of pairs distilled in parallel is
also decreased). This assumption seems reasonable, as entanglement
is, generally, the most expensive of all relevant resources, while
classical communications can be accomplished through a
conventional telephone line.

 Our investigation differs in two distinct ways from the investigations in the original
proposal for quantum repeaters \cite{qreapt}. Firstly, the focus
is shifted from reliable distribution of entanglement to resource
reduction. While in \cite{qreapt} it was shown that one could use
repeaters to prevent the exponential degradation of entanglement
with only logarithmic increase in cost,  we show that the cost of
distributing a certain fixed amount of entanglement between two
ends of a channel can be reduced using repeaters.  Secondly, we
will take a general approach based on known values and bounds on
distillable entanglement \cite{purf2,wot,ved1,ved2,rains} and
conjectured bounds \cite{conj}. In this sense, our analysis will
remain valid even if improved (or even optimal) entanglement
distillation schemes are found and even if the future advance of
technology results in completely error free quantum operations.
We should mention here that for those particular cases for which
we use known conjectures, rather than proven results, our results
may not continue to hold true if the conjecture is shown invalid.
However, exactly the same methodology as ours can still be used
to investigate resource reduction via repeaters using any
accepted nonzero lower bounds on distillable entanglement. We
should also mention that we consider only asymptotic distillation
procedures, primarily because of they give simple entropic
expressions for entanglement. This, of course, provides a
restriction to the amount by which resources can be reduced while
still keeping our analysis valid. We would require asymptotic
distillation results to remain valid even for the sets of
entangled pairs being distilled between two adjacent repeaters or
a repeater and an end of the channel. This lower number of
entangled pairs must thus, already, be very large. So we would
want the resource reduction to be not so much as to make some of
the distillation procedures nonasymptotic. For this reason, we
will calculate resource reduction as a ratio instead of as a
difference and it will remain valid as long as the initial number
$N$ of sources is so large, that even during the reduced resource
operation with repeaters, all the asymptotic results remain
applicable.

   In the next section we proceed to investigate resource
   reduction by repeaters for those channels which result in
   states whose distillable entanglement is exactly known.

\section{Exact results}

\subsection{Watched amplitude damping channel}

We will start with the watched amplitude damping channel, as the
output states of this channel are pure and we can readily use the
known asymptotic distillation results for pure states
\cite{purf1}. The amplitude damping channel has the following
effect on states $\dket{0}$ and $\dket{1}$ of a qubit
\cite{revs2,revs3}:

\[ \ba{lll}
\dket{0}\dket{0}_E & \rightarrow & \dket{0}\dket{0}_E \\
\dket{1}\dket{0}_E & \rightarrow & \sqrt{1-p}\dket{1}\dket{0}_E +
\sqrt{p}\dket{0}\dket{1}_E \ea \]
where $p$ is a parameter that is
related to the length of the channel by
\[ p = 1 - e^{-2\Gamma}, \]
in which $\Gamma$ is a quantity proportional to the length of the
channel and the subscript $E$ stands for the environmental state.
The expression for $p$ above has been so chosen that at infinite
length of the channel, the state of the system becomes
$|0\rangle$. When the environment is being watched, and found to
be in the state $|0\rangle_E$, we can consider the evolution of
the state to be given by \cite{revs3}

\[ \ba{lll}
\dket{0} & \rightarrow & \dket{0} \\
\dket{1}& \rightarrow & e^{-\Gamma}\dket{1} \\
\ea \] and this occurs for a state $ \alpha\dket{0} +
\beta\dket{1} $ with the probability $\frac{ |\alpha|^2 +
|\beta|^2 e^{-2\Gamma} }{ |\alpha|^2 + |\beta|^2}$. We consider
sending qubits in two types of initial states down the channel,
namely

\[ \rt \left( \dket{01} + \dket{10} \right) \mbox{ and } \rt \left( \dket{00} + \dket{11} \right) \]
We consider each state separately, because the second requires
purification after passing down a watched amplitude damping
channel, but the first does not.

\subsubsection{The case $ \rt \left( \dket{01} + \dket{10} \right) $}

\begin{figure}
\begin{center}
\leavevmode \epsfxsize=8cm \epsfbox{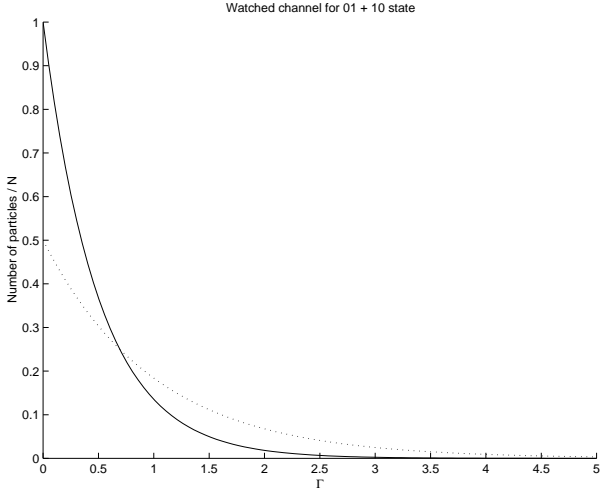}
\caption{The figure shows the variation of the number of
maximally entangled pairs established between the ends of a
watched amplitude damping channel following two different methods
of distribution. The bold line shows the case of a channel without
repeaters and the dotted line shows the case of a channel with
one repeater. In both cases, the same number of initial sources
of entanglement are used and all these sources emit the state
$|01\rangle+|10\rangle$.} \label{fig:figa}
\end{center}
\end{figure}

Initially, consider sending a state $ \rt \left( \dket{01} +
\dket{10} \right)$ down the channel with $N$ sources of entangled
qubits at the precise midpoint of the channel as shown below:

\begin{center}
\begin{picture}(200,100)(-100,-50)
\put(-90,0){\line(70,0){70}} \put(20,0){\line(70,0){70}}
\put(-100,0){2} \put(95,0){1} \put(-90,0){\circle*{3}}
\put(90,0){\circle*{3}}\put(-17.5,8){$\leftarrow N\rightarrow$}
\put(-14,-2.5){source}
\end{picture}
\vspace*{-1.5cm}
\end{center}
In the above diagram, let $1$ and $2$ symbolize two qubits from
one of the sources reaching opposite ends of the channel. If both
the qubits pass through an amplitude damping channel on their way
to the ends of the channel, the final state of the qubits and
their environment is:

\[ \ba{lll}
|\Psi\rangle_{12E_1E_2}&=&\rt ( e^{-\Gamma} \left[ \dket{01}_{12} + \dket{10}_{12} \right] \dket{00}_{E_1E_2}  \\
 &+& \sqrt{1-e^{-2\Gamma}}\dket{00}_{12}\left[ \dket{10} +
\dket{01} \right]_{E_1E_2} ) \ea \] If the environment is being
watched then there is probability $e^{-2\Gamma}$ that the
resulting state is unchanged and maximally entangled, and
$1-e^{-2\Gamma}$ that it is a (useless) product state.  In this
way we will end up with
\begin{equation}
\label{eq:eqna} K_0= Ne^{-2\Gamma}
\end{equation}
surviving entangled pairs, where the subscript $0$ has been used
to indicate that the channel has no repeaters (is undivided).
Consider now splitting the channel into two and placing half of
the sources at the midpoint of one half and the rest half of the
sources in the midpoint of the other half in the following manner:

\begin{center}
\begin{picture}(200,100)(-100,-50)
\put(-90,0){\line(35,0){35}} \put(-40,0){\line(35,0){35}}
\put(5,0){\line(35,0){35}} \put(55,0){\line(35,0){35}}

\put(-100,5){3} \put(-10,5){4} \put(5,5){2} \put(95,5){1}
\put(-90,0){\circle*{3}} \put(90,0){\circle*{3}}
\put(5,0){\circle*{3}} \put(-5,0){\circle*{3}}
\put(-65,9.5){$\leftarrow
\frac{N}{2}\rightarrow$}\put(30,9.5){$\leftarrow
\frac{N}{2}\rightarrow$} \put(-55,-2.5){src} \put(40,-2.5){src}

\end{picture}
\vspace*{-1.5cm}
\end{center}
In the above diagram, let $1$ and $2$ stand for the qubits from
any one of the sources reaching the ends of the right half-channel
and $3$ and $4$ stand for similar qubits reaching the ends of the
left half-channel. The place where $2$ and $4$ meet, namely the
midpoint of the full-channel, is the site of a quantum repeater.
If we kept the total number $N$ of sources of entangled pairs
fixed, but still large, then $N/2$ pairs are now used for each
half-channel. For each half, we now have $ \Gamma \rightarrow
\Gamma / 2 $.  This means the number of maximally entangled pairs
available finally in each half is the product of $N/2$ multiplied
by the probability $\frac{1}{2}e^{-2\Gamma/2}$ of getting a
maximally entangled state
 {\em i.e.} $\frac{N}{2} \times e^{-2\Gamma/2}$. Let particle pairs $(1,2)$ and
$(3,4)$ in the above diagram now be taken to symbolize any two of
the several maximally entangled pairs obtained in the
half-channels. If joint projections are now performed on
particles 2 and 4 by an operator at the repeater, a maximally
entangled pair is obtained across the full channel. The number of
maximally entangled pairs obtained finally across the whole
channel is thus equal to the number of maximally entangled pairs
obtained in each half-channel prior to the joint projections on
$2$ and $4$ at the repeater. It is therefore equal to
\begin{equation}
\label{eq:eqnb}
K_1=\frac{N}{2} \times e^{-2\Gamma/2},
\end{equation}
where the subscript $1$ denotes the fact that one repeater is now
being used. So, whilst the exponent of $K_1$ is smaller compared
to that of $K_0$, there is an extra factor of $1/2$ in front of
$K_1$ compared to $K_0$. So is $K_0$ larger or smaller than $K_1$?
Fig.\ref{fig:figa} plots $K_0$ and $K_1$ as functions of $\Gamma$.
This shows that after a certain value of $\Gamma$, it is better
(in terms of obtaining higher final entanglement) to split the
full-channel to two half-channels with a repeater in the middle.

    This idea can be generalized to splitting the channel into $m$
sections.  Eq.(\ref{eq:eqnb}) can then be generalized to:
\begin{equation}
\label{eq:watchm} K_{m-1}=\frac{N}{m} \times e^{-2\Gamma/m}
\end{equation}
In each case, after a certain value of $\Gamma$, it becomes better
to use 3 rather than 2 or 4 rather than 3 etc. sections of the
channel in order to obtain maximum entanglement.

\begin{figure}
\begin{center}
\leavevmode \epsfxsize=8cm \epsfbox{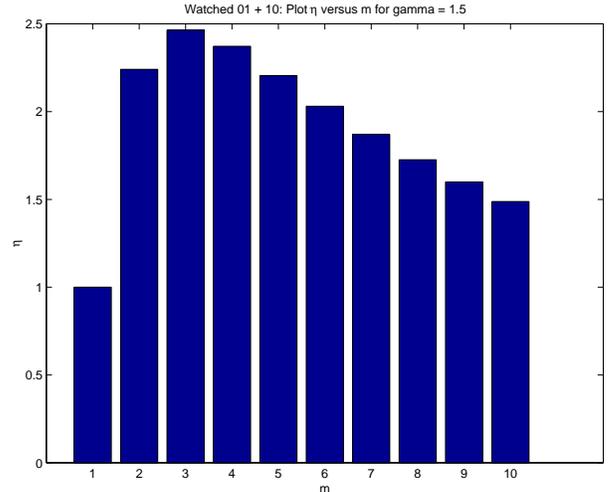}
\caption{The resource reduction ratio $\eta$ for a watched
amplitude damping channel is shown for various values of the
number $m$ of sections of the channel (when repeaters are used).
The state emitted by the sources is taken to be
$|01\rangle+|10\rangle$ and $\Gamma=1.5$.} \label{fig:b}
\end{center}
\end{figure}

  We now show how this idea can be used to achieve a
resource reduction in entanglement distribution via repeaters
({\em i.e.}, via splitting of a channel into several segments). We
will show that we will need a smaller number of sources of
entangled pairs to obtain the same amount of final entanglement
across two ends of the channel. Consider relabeling $N$ in
Eq.(\ref{eq:watchm}) to $N'$ and setting that expression equal to
$K_0$. This condition demands that with $m-1$ repeaters and
$N^{'}$ sources of entanglement we produce the same number of
maximally entangled pairs finally across the channel as the
undivided channel with $N$ sources placed in the middle. We can
then compare the ratio $N/N'$, that is, the ratio of input pairs
(sources) required for the undivided channel to the number of
input pairs (sources) required for the channel with repeater
stations. Let us label this ratio $\eta$. Then we have
\begin{equation}
\eta = \frac{e^{-2\Gamma/m}}{m \times e^{-2\Gamma}}
\end{equation}
When $\eta$ is greater than one, it means the channel with
repeaters needs fewer input pairs (sources) i.e. less resources.
Graph \ref{fig:b} plots $\eta$ versus $m$ - the number of
sections in a channel with repeaters, for a fixed value of
$\Gamma=1.5$. At $m=3$ sections (or two repeaters), the resource
reduction over the undivided channel is highest. Two repeaters is
thus the the optimal way to distribute entanglement for
$\Gamma=1.5$. In this case, using just $\frac{2}{5}$th of the
amount of initial resources, we achieve the same output of
entanglement as an undivided channel.

\begin{figure}
\begin{center}
\leavevmode \epsfxsize=8cm \epsfbox{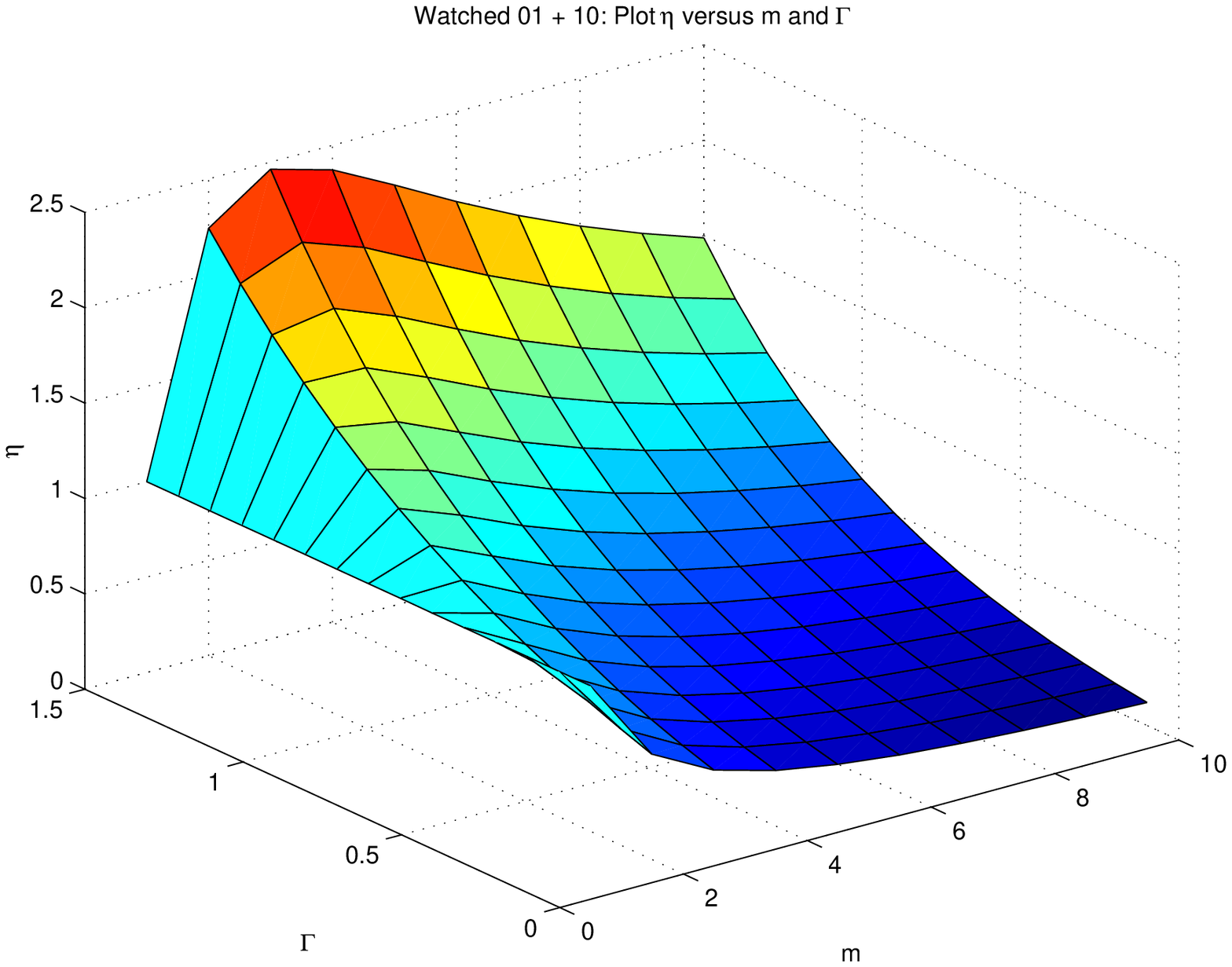} \caption{The
resource reduction ratio $\eta$ for a watched amplitude damping
channel is shown for various values of the number $m$ of sections
of the channel (when repeaters are used) and various values of
$\Gamma$. The state emitted by the sources is taken to be
$|01\rangle+|10\rangle$. The plot illustrates the fact that if
the number of repeaters is kept fixed, more and more resource
reduction is obtained with increase of the channel length.}
\label{fig:w3d}
\end{center}
\end{figure}

Fig.\ref{fig:w3d} is a plot in 3D, extending Fig \ref{fig:b} over
a range of $\Gamma$ to show the variation of the optimal number
of divisions of the channel with $\Gamma$. We see that as $\Gamma$
(which is proportional to the channel length) increases, it
becomes more and more advantageous in terms of resource reduction
to use repeaters.

\subsubsection{The case $ \rt \left( \dket{00} + \dket{11} \right) $}

We now consider sending two qubits in the initial state $ \rt
\left( \dket{00} + \dket{11} \right) $ down an amplitude-damped
channel.  The resulting state is:

\[ \ba{lll}  |\Psi^{'}\rangle_{12E_1E_2}&=&\rt( \sqrt{ 1 + e^{-4\Gamma}}[ \frac{ \dket{00} + e^{-2\Gamma} \dket{11}}{ \sqrt{ 1 + e^{-4\Gamma}} }
]_{12} \dket{00}_{E_1E_2} \\ &+& e^{-\Gamma}\sqrt{1-e^{-2\Gamma}}
\dket{01}_{12}\dket{10}_{E_1E_2} \

\\
&+& e^{-\Gamma}\sqrt{1-e^{-2\Gamma}}
\dket{10}_{12}\dket{01}_{E_1E_2} \\&+& (1-e^{2-\Gamma})
\dket{00}_{12}\dket{11}_{E_1E_2}) \ea  \]
 Then,
if the environment is being monitored, there is a probability of
$\frac{1}{2} \left( 1+e^{-4\Gamma}\right)$ that the state
observed is (corresponds to the state $\dket{00}_{E_1E_2}$ of the
environment)

\begin{equation}
|\Psi^{'}_c\rangle_{12}=\frac{1}{ \sqrt{1+e^{-4\Gamma}} } \left(
\dket{00} + e^{-2\Gamma} \dket{11} \right) \label{eq:c}
\end{equation}
where the subscript $c$ represents the fact that this state is a
result of conditional evolution. $|\Psi^{'}_c\rangle_{12}$ is not
maximally entangled and must now be purified.  When a large
number of pure states like these are distilled, the number of
maximally entangled states that can be produced is given by  $N
S(\rho^{r})$ \cite{purf1}, where $\rho^{r}$ is the reduced
density matrix
 of any one of the qubits and $S(\rho)$ is the von-Neumann entropy
 given by $-\mbox{Tr}\rho \log_{2}{\rho}$.

If we began with an undivided channel and $N$ pairs were
dispatched from $N$ sources, all placed at the midpoint, then $
\frac{1}{2} \left( 1+e^{-4\Gamma}\right)N$ reach opposite ends of
the channel in the state $|\Psi^{'}_c\rangle_{12}$ and rest reach
the ends disentangled. One can locally purify these states to
obtain

\[ K^{'}_0=\frac{1}{2} \left( 1+e^{-4\Gamma} \right) S(\rho^{r}) N \]
maximally entangled pairs between the two ends of the channel.

 Now consider splitting this channel into two halves as before. At
the midpoint of each half we place $N/2$ sources and make them
emit maximally entangled pairs. The probability that an entangled
state arrives at the ends of a half-channel is $\frac{1}{2} (
1+e^{-4\Gamma/2})$.  Each half then purifies its pairs
(independently of the other half). So the number of maximally
entangled pairs in each half, after purification is

\[ K^{'}_1=\frac{1}{2} \left( 1+e^{-4\Gamma/2} \right) S(\rho^r_{\frac{1}{2}}) \times \frac{N}{2}, \]
where $\rho^r_{\frac{1}{2}}$ is the reduced density matrix of one
of the qubits in the state shared between any end and the repeater
before purification. The subscript $1/2$ in
$\rho^r_{\frac{1}{2}}$ has been used to indicate the fact that
now each qubit has traveled half the distance it would have had
to travel for an undivided channel. The disentangled pairs after
purification are discarded and only the pairs in a maximally
entangled state are kept. A projection by a central operator to
connect up the ends of the full channel with one maximally
entangled state now requires one pair from each half, so
$K^{'}_1$ is also the number of maximally entangled pairs that
can be produced between the two ends of the entire channel.

\begin{figure}
\begin{center}
\leavevmode \epsfxsize=8cm \epsfbox{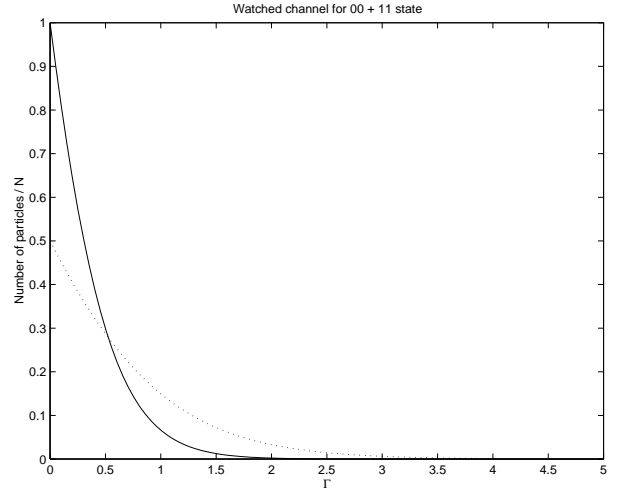}
\caption{The figure shows the variation of the number of
maximally entangled pairs established between the ends of a
watched amplitude damping channel following two different methods
of distribution. The bold line shows the case of a channel without
repeaters and the dotted line shows the case of a channel with
one repeater. In both cases, the same number of initial sources
of entanglement are used and all these sources emit the state
$|00\rangle+|11\rangle$.} \label{fig:c}
\end{center}
\end{figure}

The number of maximally entangled states $K^{'}_0$ produced by an
undivided channel and $K^{'}_1$ produced due to two half-channels
is plotted in Fig.\ref{fig:c}. Analogous to the case of $\dket{01}
+ \dket{10}$, at a certain value of $\Gamma$, we get more final
entanglement between the two ends of the channel if we split the
channel and place repeaters. When these results are compared to
distributing the state $ \dket{01} + \dket{10} $ we see that they
are not as good - i.e. using $ \dket{01} + \dket{10} $ will give
more maximally entangled states across the full channel.

\begin{figure}
\begin{center}
\leavevmode \epsfxsize=8cm \epsfbox{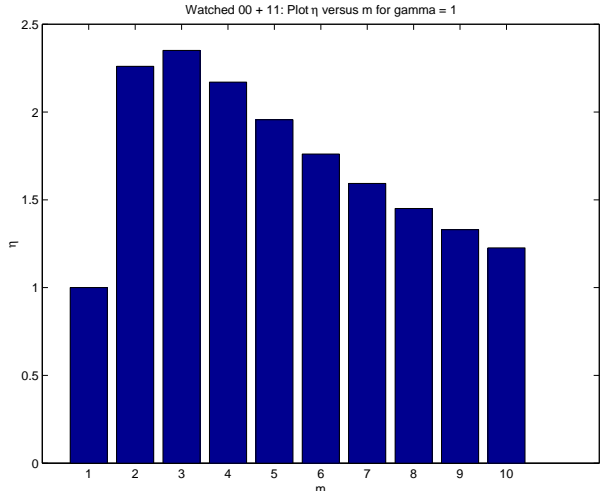}
\caption{The resource reduction ratio $\eta$ for a watched
amplitude damping channel is shown for various values of the
number $m$ of sections of the channel (when repeaters are used).
The state emitted by the sources is taken to be
$|00\rangle+|11\rangle$ and $\Gamma=1$.} \label{fig:d}
\end{center}
\end{figure}

As before, this idea can be generalized to splitting the channel
into $m$ sections.  Each value of $m$ becomes the optimal number
of channel sections (for obtaining the maximum final entanglement,
if the initial resources were kept constant) after a certain
value of $\Gamma$. Again, as before, we can calculate $\eta$, the
ratio of the number of initial sources required for the undivided
channel to the number of initial sources required for the channel
with repeaters. Fig.\ref{fig:d} plots $\eta$ versus $m$ - the
number of sections in the channel with repeaters, for $\Gamma=1$.
We see that in this case, $m=3$ sections (two repeaters), gives
the largest resource reduction over the undivided channel.

\subsection{Bit-flip channel}

\begin{figure}
\begin{center}
\leavevmode \epsfxsize=8cm \epsfbox{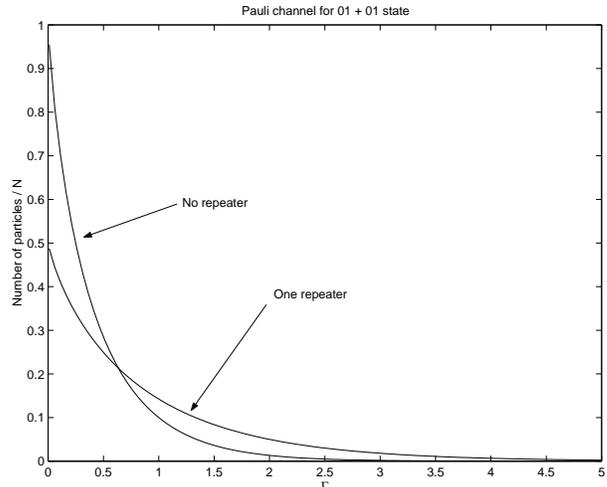} \caption{The
figure shows the variation of the number of maximally entangled
pairs established between the ends of a bit-flip channel following
two different methods of distribution: without repeaters and with
one repeater. In both cases, the same number of initial sources
of maximally entangled pairs are used.} \label{fig:pauli}
\end{center}
\end{figure}

We now consider sending the states through a different kind of
channel - a Pauli channel where there is just one type of error.
The reason for this is the fact that the final states produced on
sending a pair of qubits in a maximally entangled initial state
through this channel is a mixture of two Bell states. The
distillable entanglement for such states is exactly known
\cite{ved2,rains}. If we take the single error in the channel to
be of the "bit-flip" type (called the bit-flip channel
\cite{revs2}), an initial state
$|\psi^{+}\rangle=\frac{1}{\sqrt{2}}(\dket{01} + \dket{10})$
transforms as

\[
\dket{\psi^+}\dbra{\psi^+} \rightarrow \lambda
\dket{\psi^+}\dbra{\psi^+} + (1 - \lambda)
\dket{\phi^+}\dbra{\phi^+},
\]
where $|\phi^{+}\rangle=\frac{1}{\sqrt{2}}(\dket{00} +
\dket{11})$. We relate the parameter $\lambda$ to the length of
the channel (which is proportional to $\Gamma$) by

\[ \lambda \equiv \frac{1 + e^{-\Gamma}}{2} .\]
From Ref.\cite{ved2,rains} we know that the distillable
entanglement for the state $\Lambda=\lambda
\dket{\psi^+}\dbra{\psi^+} + (1 - \lambda)
\dket{\phi^+}\dbra{\phi^+}$ is given by $1-S(\Lambda)$, where
$S(\Lambda)$ denotes the von Neumann entropy of the state
$\Lambda$. On using an undivided channel with $N$ sources placed
at the midpoint, we would obtain $K^{''}_0=N[1-S(\Lambda))]$
particle pairs in a final maximally entangled state across the
full length of the channel. On the other hand, if we divided the
channel into $m$ sections, and placed $N/m$ sources at the
midpoint of each segment, the final number of maximally entangled
pairs across the full channel will be
$K^{''}_m=\frac{N}{m}[1-S(\Lambda_m))]$ where $\Lambda_m=\lambda_m
\dket{\psi^+}\dbra{\psi^+} + (1 - \lambda_m)
\dket{\phi^+}\dbra{\phi^+}$, in which $\lambda_m=\frac{1 +
e^{-\Gamma/m}}{2}$. In Fig.\ref{fig:pauli}, we have plotted
$K^{''}_0$ and $K^{''}_1$ as a function of $\Gamma$. It is clear
that for $\Gamma$ exceeding a certain value, more final
entanglement between the ends of the channel for a given initial
resource  is obtained on dividing the channel into two parts and
using a repeater at their junction.

\begin{figure}
\begin{center}
\leavevmode \epsfxsize=8cm \epsfbox{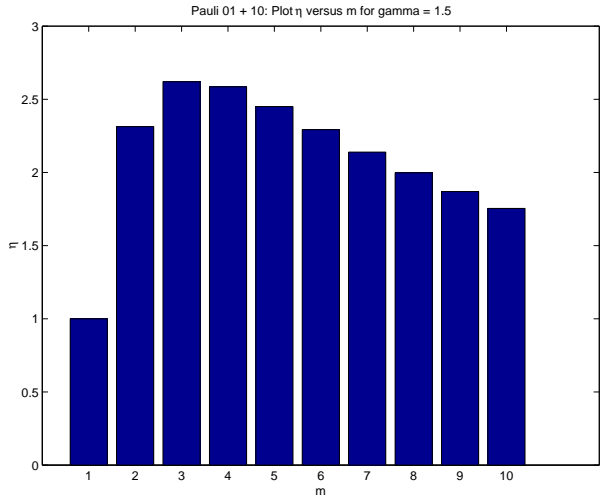} \caption{The
resource reduction ratio $\eta$ for a bit-flip channel is shown
for various values of the number $m$ of sections of the channel
(when repeaters are used). Here $\Gamma=1.5$.}
\label{fig:paulired}
\end{center}
\end{figure}

As in the previous cases, we now consider how repeaters can be
used to achieve the same final entanglement using less initial
resources. The ratio of the number of initial sources required
for the undivided channel to that required when the channel is
split into $m$ sections with repeaters, is given by

\[
\eta = \frac{1 - S\left(\Lambda_m\right)}{m (1 - S(\Lambda)) }
\]
This ratio $\eta$ is plotted in Fig (\ref{fig:paulired}) for
$\Gamma = 1.5$  and it can be seen that use of $1$ repeater to $9$
repeaters allow a smaller initial resource to create the same
final amount of entanglement between the two ends of the full
channel.

Exactly the same results on resource reduction would hold for
phase flip \cite{revs3} and bit-phase flip \cite{revs3} channels,
as they all take initially maximally entangled states to final
states which are mixtures of two Bell states.

\section{Conjecture based results}

We now consider more general channels. However, the final states
produced when maximally entangled qubit pairs pass through these
channels are such that their distillable entanglement is not
known. For such channels we will use an established upper bound
and a conjectured lower bound on distillable entanglement for
comparing distribution with and without repeaters.

  The well established upper bound on distillable entanglement $E_D$
is the entanglement of formation $E_F$ \cite{purf2,wot}, which is
also simple to calculate for a $2\times 2$ system \cite{wot}. As
a lower bound on distillable entanglement, we will use a
conjecture recently made by Horodecki, Horodecki and Horodecki
\cite{conj}. This gives a lower bound on the distillable
entanglement of a state $\rho_{12}$ of qubits $1$ and $2$ by

\[ E_D \geq \left.
\ba{ll} S(\rho_1) - S(\rho_{12})
\\
S(\rho_2) - S(\rho_{12}) \ea \right\} \mbox{ whichever is lower},
\]
where $\rho_1$ and $\rho_2$ denote the reduced density matrix of
any one of the qubits. In Ref.\cite{conj}, this conjecture has
been used to investigate an unified approach to quantum channel
capacities and there is a significant body of evidence which lends
support to this conjecture \cite{support}.

  For each type of channel, we first find the final density matrix
$\rho(\Gamma)$ of two qubits for distribution through an undivided
channel (with the source placed at the midpoint). We then
calculate the density matrix $\rho(\Gamma/m)$ of entangled qubits
reaching adjacent repeaters from the midpoints of the
corresponding sections when the channel is split into $m$
sections. We now compare the lower bound $S(\rho_1) -
S(\rho_{12})$ or $S(\rho_2) - S(\rho_{12})$ on distillable
entanglement for the channel with repeaters with the upper bound
$E_F$ for an undivided channel. In this way we ensure that we are
imparting no intrinsic advantage to the case of repeaters over the
case of an undivided channel. As before, the resource reduction is
quantified by the ratio $\eta$ (defined in the same manner as
earlier) and this ratio is given (assuming, without loss of
generality, $S(\rho_1)$ to be the smaller among $S(\rho_1)$ and
$S(\rho_2)$) by

\[
\eta = \frac{S \left( \rho_{1} \left( \Gamma /m\right) \right) -
S \left( \rho_{12} \left( \Gamma / m \right) \right) }{m E_F
\left( \rho (\Gamma) \right)}.  \] We consider three types of
channels
 \cite{revs2,revs3}
\begin{itemize}
\item Amplitude damping channel - i.e.\ the 'full' version of the watched channel
\item Depolarising channel
\item Phase damping channel
\end{itemize}

\begin{figure}
\begin{center}
\leavevmode \epsfxsize=8cm \epsfbox{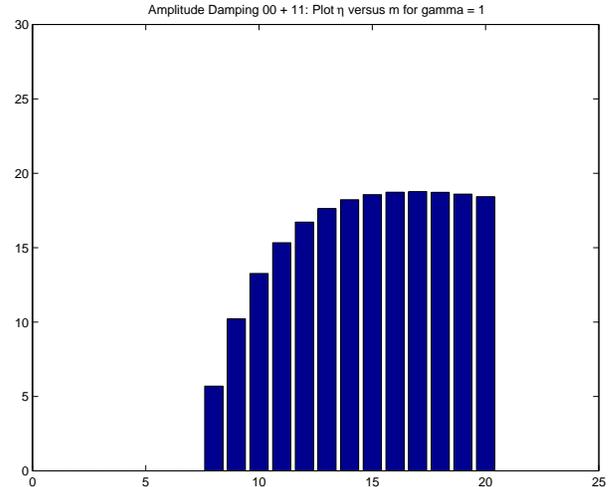} \caption{The
resource reduction ratio $\eta$ for an amplitude damping channel
is shown for various values of the number $m$ of sections of the
channel (when repeaters are used). The state emitted by the
sources is taken to be $|00\rangle+|11\rangle$ and $\Gamma=1$.
Before dividing the channel to $8$ sections, no entanglement at
all can be established between the ends of the channel. We see
that $\eta$ increases at first with $m$ and then starts to
decrease.} \label{fig:amp00}
\end{center}
\end{figure}

\subsection{Amplitude damping channel}

\begin{figure}
\begin{center}
\leavevmode \epsfxsize=8cm \epsfbox{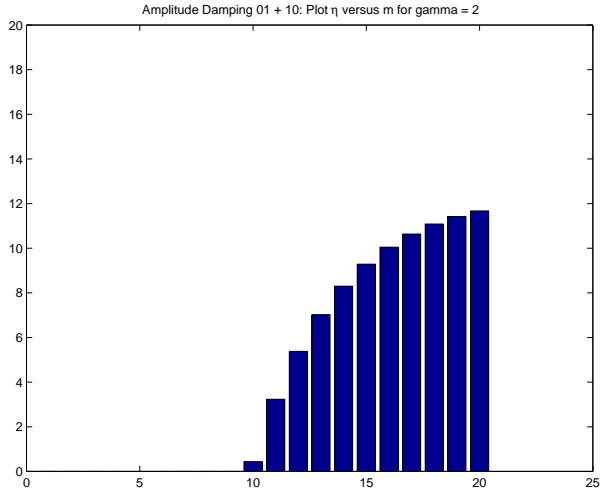} \caption{The
resource reduction ratio $\eta$ for an amplitude damping channel
is shown for various values of the number $m$ of sections of the
channel (when repeaters are used). The state emitted by the
sources is taken to be $|01\rangle+|10\rangle$ and $\Gamma=2$.
Before dividing the channel to $10$ sections, no entanglement at
all can be established between the ends of the channel and only
from $11$ sections onwards we see a resource reduction.}
\label{fig:amp01}
\end{center}
\end{figure}

For the amplitude damping channel we consider sending two states
down the channel, $ \rt(\dket{01} + \dket{10}) $ and $ \rt
(\dket{00} + \dket{11}) $.  This is because of the fact that the
amplitude damping channel affects only the $ \dket{1} $ state and
therefore has quite different effects on the above two states. The
amplitude damping channel affects states in the following way
\cite{revs2,revs3}

\[ \ba{lll}
\dket{0}_A\dket{0}_E & \rightarrow & \dket{0}_A\dket{0}_E \\
\dket{1}_A\dket{0}_E & \rightarrow & \sqrt{1-p}
\dket{1}_A\dket{0}_E + \sqrt{p} \dket{0}_A\dket{1}_E \ea .\] We
relate $p$ to $\Gamma$ by the following
\[ p = 1 - e^{-2\Gamma} \]
Fig.\ref{fig:amp01} shows $\eta$ for $ \rt(\dket{01} + \dket{10})
$ and Fig.\ref{fig:amp00} for $ \rt (\dket{00} + \dket{11}) $.

We see that in the case of using the state $\rt(\dket{01} +
\dket{10})$, there could almost be a $20$ fold reduction of
resources and in the case of the state $ \rt (\dket{00} +
\dket{11}) $, there can be nearly a $12$ fold reduction of
resources if repeaters are used.
\subsection{Phase damping channel}

\begin{figure}
\begin{center}
\leavevmode \epsfxsize=8cm \epsfbox{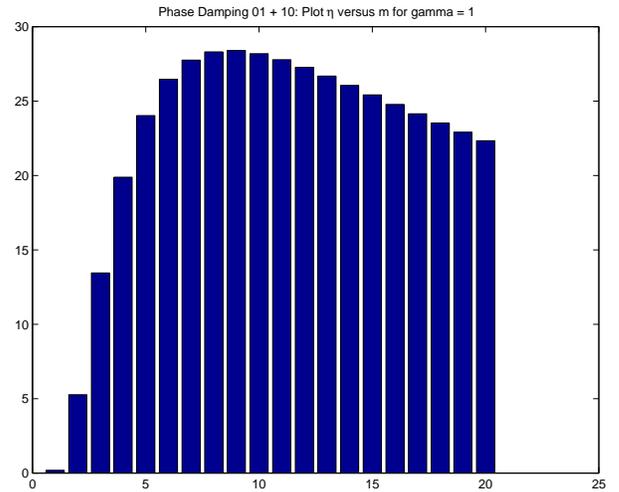} \caption{The
resource reduction ratio $\eta$ for a phase damping channel is
shown for various values of the number $m$ of sections of the
channel (when repeaters are used). Here $\Gamma=1$. As soon as the
channel is split to $2$ sections, there is a resource reduction
and this first increases with $m$ and then starts decreasing
again.} \label{fig:phase}
\end{center}
\end{figure}

The phase damping channel affects states as \cite{revs2,revs3}

\[ \ba{lll}
\dket{0}_A\dket{0}_E & \rightarrow & \sqrt{1-p} \dket{0}_A\dket{0}_E + \sqrt{p} \dket{0}_A\dket{1}_E \\
\dket{1}_A\dket{0}_E & \rightarrow & \sqrt{1-p} \dket{1}_A\dket{0}_E + \sqrt{p} \dket{1}_A\dket{2}_E
\ea \]
We relate $p$ to $\Gamma$ by the following:
\[ p \rightarrow 1 - e^{-\Gamma} \]
This dependence of $p$ on $\Gamma$ has been so chosen that the
entanglement of qubits propagating in this channel completely
vanishes at $\Gamma\rightarrow \infty$. Fig.\ref{fig:phase} shows
$\eta$ for this channel.

In this case we see that the use of repeaters can lead to almost
up to $28$ fold reduction in resources.

\begin{figure}
\begin{center}
\leavevmode \epsfxsize=8cm \epsfbox{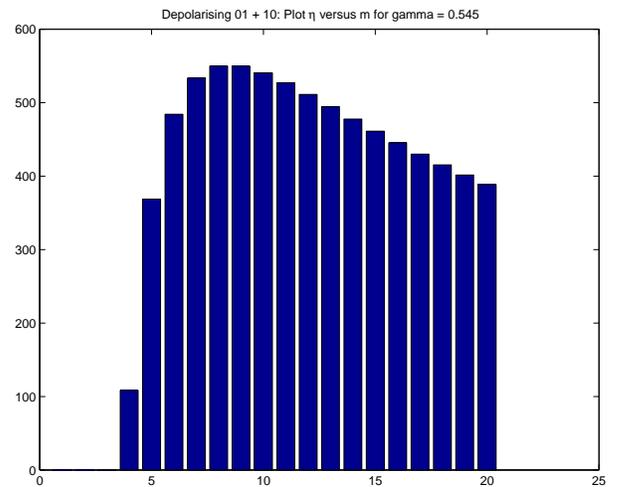} \caption{The
resource reduction ratio $\eta$ for a depolarizing channel is
shown for various values of the number $m$ of sections of the
channel (when repeaters are used). Here $\Gamma=0.545$. Before
dividing up the channel to $4$ sections, no entanglement can be
established between its ends. The resource reduction ration
$\eta$ initially increases with $m$ and then starts decreasing
again.} \label{fig:depol}
\end{center}
\end{figure}

\subsection{Depolarizing channel}

The depolarizing channel affects states as

\[ \rho \rightarrow \rho' = (1-p)\rho + \frac{p}{3} \left( \sigma_1\rho\sigma_1 + \sigma_2\rho\sigma_2 + \sigma_3\rho\sigma_3 \right) \]
We relate $p$ to $\Gamma$ by the following:
\[ p \rightarrow \frac{3(1-e^{-\Gamma})}{4} \]
This dependence of $p$ on $\Gamma$ has been so chosen that the
entanglement of qubits propagating in this channel completely
vanishes at $\Gamma\rightarrow \infty$. Fig.\ref{fig:depol} shows
$\eta$ for this channel.

In this case we see that use of repeaters can lead to, in the
best case, a reduction as large as $500-600$ fold in resources.
In the case of all channels, we see that the degree of resource
reduction increases at first with the number of repeater stations
and then starts to decrease on further increase of the number of
repeaters.

\section{Conclusions}
We have shown that the amount of initial resource of entanglement
needed to establish a certain final amount of entanglement between
two ends of a noisy channel can be reduced, for certain lengths
of the channel, by using quantum repeaters. This result has
important bearing on cost minimization in entanglement
distribution. We have used a variety of channels and asymptotic
values and bounds on distillable entanglement to arrive at our
results. While in the original papers on quantum repeaters
\cite{qreapt}, the emphasis was on preventing the exponential
degradation of entanglement through practically motivated
specific purification procedures, our emphasis is more on
attempting to use general results on distillable entanglement. As
such, our results are quite independent of the type of the
purification procedures used. We also use asymptotic measures of
entanglement \cite{purf2,wot,ved1,ved2} in our analysis in
contrast to the earlier treatment \cite{qreapt} based on fidelity
of transmission of the states. Though the field of entanglement
measures is very well developed, there have not been many
attempts (apart from the natural application to entanglement
purification \cite{ved2}) to link these measures to practical
issues such as the cost of entanglement distribution. We regard
our analysis to be a step in this direction. All our analysis has
been based on asymptotic purification procedures in order to use
simple entropic quantities as bounds on distillable entanglement.
However, the field of local entanglement manipulations for
non-asymptotic cases has developed recently \cite{finite}. It
would be interesting to explore the possibility of resource
reduction by repeaters in the case when only a small number of
entangled qubit pairs are transmitted through the channel at a
time.

AH thanks the UK EPSRC (Engineering and Physical Sciences
Research Council) for financial support.

\end{multicols}

\begin{multicols}{2}
\end{multicols}
\end{document}